\documentstyle[aps,12pt,tighten]{revtex}
\begin{document}
\draft
\input{epsf}

\title{ Helicity Zero Particles}

\author{W. A. Perkins}

\address{Perkins Advanced Computing Systems,\\ 12303 Hidden Meadows
Circle, Auburn, CA 95603, USA\\E-mail: wperkins@aub.com} 

\maketitle

\begin{abstract}
In this paper we consider the possibility that a vector particle with mass might exist in only one helicity state, rather than the usual three states with helicity equal to +1, -1, and 0. Massless particles, of course, need only have one helicity state. (For invariance under parity, they need two.) We show that a massive vector particle can exist only in the helicity-0 state, if it is composed of a fermion-antifermion pair and they are massless. This requires the mass to be generated by the interaction between the massless particles. An interaction of the form $\Psi^\dagger i \gamma_4 \gamma_{\mu} \Psi$ is attractive between particle and antiparticle and preserves helicity. Methods for experimentally distinguishing an helicity-0 vector particle from both a spin-0 pseudoscalar particle and a spin-1 vector particle are discussed. 
\end{abstract}


\pacs{PACS numbers: 12.60.Rc}

\section{Introduction}
\label{sec.intro}

The purpose of this paper is to investigate the possibility of a massive vector particle having less than all three helicity states. We are all familiar with massless particles that only exist in two helicity states (photon) and one helicity state (two-component neutrino). Here we will attempt to formulate a theory of a massive vector particle that only exist in the helicity-0 state. In Sec.~\ref{sec.concl} we will discuss a method for distinguishing an helicity-0 vector particle from a pseudoscalar particle or from a spin-1 vector particle.

Our definition of a vector particle is: A vector particle is a particle whose field transforms as a four-vector under Lorentz transformations. Notice that there is no mention of spin in the definition. As Veltman noted (see Ref.~\cite{veltman}, p.~169), ``We can freely use scalar fields, vector fields, spinor fields, as long as the theory gives rise to results agreeing with the observed data. One of the required properties is Lorentz invariance, and we must take care that Lorentz invariance is maintained. Our classification in terms of the fields mentioned is really done that way to keep this invariance transparent.''

To determine the spin corresponding to a given field requires some assumptions and a theoretical calculation. The result, as given by Roman (see Ref.~\cite{roman}, p.~99), is, ``in the case of a scalar field ($j=j \prime = 0$) we need no supplementary condition; the field has the unique spin value zero. The same holds for the elementary spinor representations ${\cal D}^{{1 \over 2} 0}$ and ${\cal D}^{0 {1 \over 2}}$ (and also their direct sum), which give the unique spin value ${1 \over 2}$. However, for a four-vector field, which belongs to the ${\cal D}^{{1 \over 2} {1 \over 2}}$ representation, $4 j j \prime = 1$; thus {\it one} supplementary condition is required in order to have the unique spin value 1 and to exclude the value zero.'' This supplementary condition is the well-known Lorentz condition and applies to the photon field. However, there are four-vector fields for which the Lorentz condition does not apply as illustrated near the end of Sec.~\ref{sec.polar}.

The two obvious kinematic properties of elementary particles are spin and mass, and they should be describable by quantities that are invariant under relativistic transformations. As Wigner first showed~\cite{wigner1} the relevant group in considering the spin of a particle is the Poincare group as spin is an invariant under that group.

Since the Poincare group is of rank 2, there are just two Casimir invariants~\cite{tung},
\begin{eqnarray}     
C_1 = P_{\mu} P_{\mu}, \nonumber\\ 
C_2 = W_{\mu} W_{\mu}, 
\label{eqn.01}
\end{eqnarray} 
where $W_{\mu}$ is the Pauli-Lubanski vector and is defined by, 
\begin{equation}     
W_{\mu} = {1 \over 2} \: \epsilon_{\mu \nu \rho \sigma} J_{\nu \rho} P_{\sigma}, 
\label{eqn.02}
\end{equation} 
and $ P_{\mu}$ is the momentum operator and the $J_{\nu \rho}$'s are the generators for Lorentz transformations. The representations of the Poincare group depend upon whether $M > 0$ or $M = 0$, where $M$ the mass of the particle.

For $M > 0$, the momentum is time-like, and the little group (i.e. the subgroup of the Poincare group that leaves $p_{\mu}$ invariant) is the group of rotations. We will represent a state by $| {\bf p} \lambda>$, where 
${\bf p}$ is the momentum and $\lambda$ is the helicity. For a rest state, 
${\bf p} = {\bf p_0}$,    
\begin{eqnarray}     
P_{\mu} | {\bf p_0} \lambda> \; = \; p^{\: r}_{\mu} | {\bf p_0} \lambda>, \nonumber\\ 
J^2 |{\bf p_0} \lambda> \; = \;  s(s+1) | {\bf p_0} \lambda>, \nonumber\\ 
J_3 | {\bf p_0} \lambda> \; = \;  \lambda | {\bf p_0} \lambda>, 
\label{eqn.03}
\end{eqnarray} 
where $p^{\: r}_{\mu} = (0,0,0,iM)$ and $s$ is the intrinsic spin.

The helicity is transformed among the $2s+1$ possible states for this massive case. The eigenvalue of the first Casimir operator, $c_1 = -M^2$, and the eigenvalue of the second Casimir operator, $c_2 = -M^2 s(s + 1)$. 

For $M = 0$, the momentum is light-like, and spin is no longer described by the rotational group. The little group in this case is a combinations of boosts and rotations that surprisingly depend upon the boost parameters. The state of a massless particle is characterized by the helicity, $\lambda$. Unlike the massive case, the helicity is invariant under Lorentz transformations. For a particle moving along the third axis,

\begin{eqnarray}     
P_{\mu} | {\bf p^{\ell}} \: \lambda> \; = \;  p^{\ell}_{\mu} | {\bf p^{\ell}} \: \lambda>, \nonumber\\ 
J_3 | {\bf p^{\ell}} \: \lambda> \; = \;  \lambda | {\bf p^{\ell}} \: \lambda>, \nonumber\\ 
W_i | {\bf p^{\ell}} \: \lambda> \; = \;  0, 
\label{eqn.04}
\end{eqnarray} 
where $p^{\ell}_{\mu} = (0,0,p,ip)$ and $i = 1, 2$.
Here the eigenvalues of the Casimir operators, $c_1 = 0$, and $c_2 = 0$. 

We are interested is seeing if a massive particle can exist in only one helicity state as massless particles can. However, a particle with mass can be transformed to a coordinate system in which it is at rest. In that system there is no helicity axis unless the particle has some internal structure. 

Suppose this special massive particle is composed of two spin-${1 \over 2}$ particles with the same mass, but oppositely directed helicities. We can visualize the different types of particles as shown in Fig.~\ref{f1}. In the coordinate system for which the massive particle is at rest, the two fermions would have momentum, ${\bf p}$ and ${\bf -p}$. Thus, the opposing helicities form an axis, even though the massive composite particle is as rest. If the massive particle is given a boost along any axis other than the one formed by the opposing helicities, the internal axis will not coincide with the momentum of the massive particle.


\epsfxsize=5.0in \epsfbox{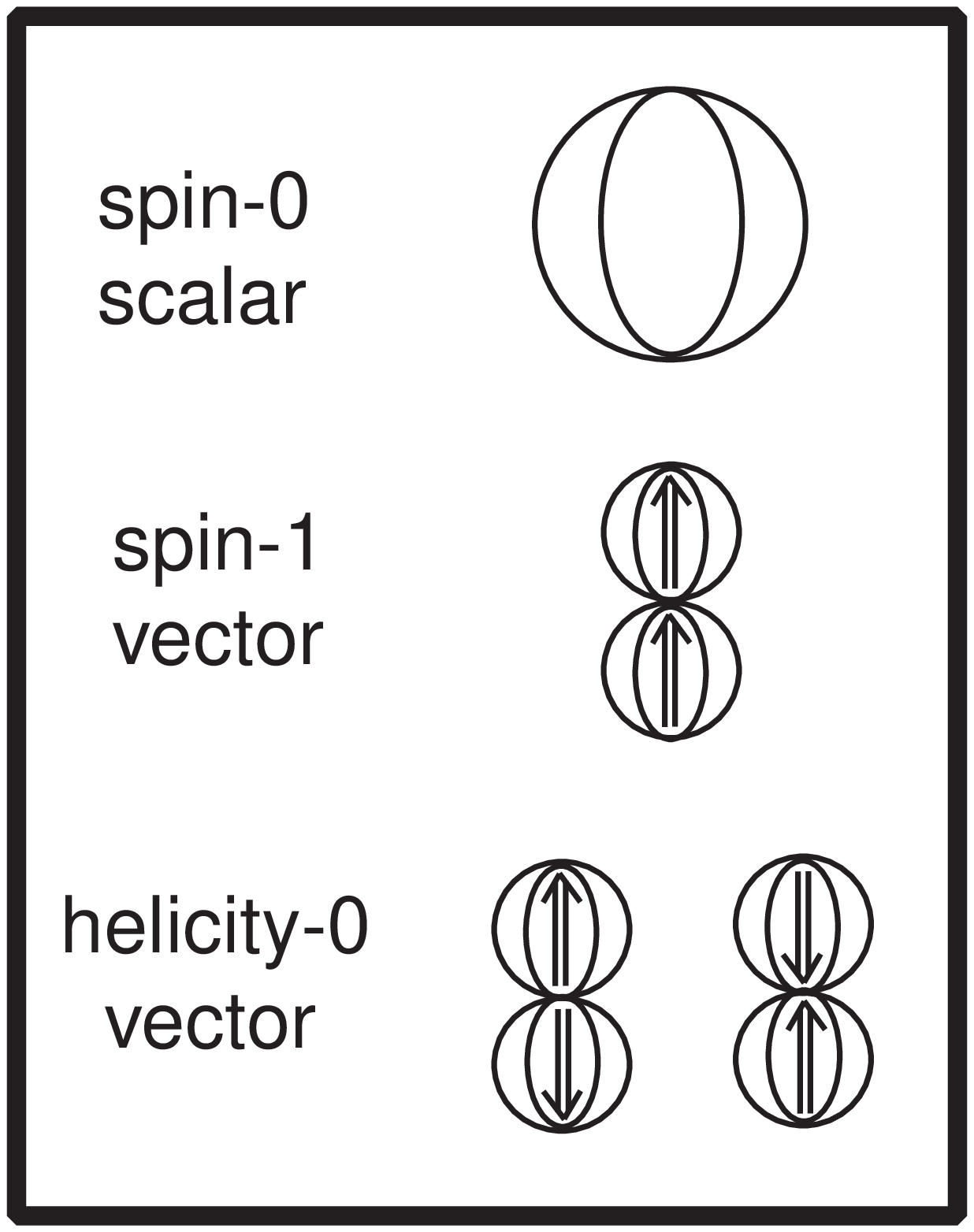}

\begin{figure}
\caption{Illustration of spin-0 scalar particle which is spherically symmetric, spin-1 vector particle formed of two spin-1/2 particles, and helicity-0 vector particle formed of two spin-1/2 particles with its two states. The arrows indicate the fermion spin directions.}
\label{f1}
\end{figure}

Let us assume this special massive particle is in the helicity-0 state given by
$(\uparrow \downarrow + \downarrow \uparrow)$, where we have used up and down arrows to denote the z-component of spin for the spin-${1 \over 2}$ particles. Normally, we would assume that this state could be transformed into the other helicity $\pm 1$ states. However, if the two fermions (that compose the massive particle) are massless, this helicity state will be just as invariant as the helicity states of the photon and neutrino. 
Mechanisms for generating mass have been suggested~\cite{nambu,kane}, and  glueballs~\cite{morningstar},  in quantum chromodynamics are an example of massive particles created by combining massless particles. 

The usual argument that a massive particle must have three polarization states does not consider the possibilities allowed by composite particles. For example, with a vector particle of mass $M$ at rest, the polarization states are (see Ref.~\cite{veltman}, p.~176) described by the three four-vectors, $e^j_{\mu}$,
\begin{eqnarray}     
e^1_{\mu} = {1 \over \sqrt{2}}(1,0,0,0), \nonumber\\ 
e^2_{\mu} = {1 \over \sqrt{2}}(0,1,0,0), \nonumber\\ 
e^3_{\mu} = {1 \over \sqrt{2}}(0,0,1,0).  
\label{eqn1}
\end{eqnarray} 

By rotating about a suitable axis, any of these can be transformed into the others. For example, if we rotate about the first axis,
\begin{eqnarray}
{1 \over \sqrt{2}} \left( \begin{array}{cccc}
1 & 0 & 0 & 0 \\
0 & \cos\theta & \sin\theta & 0 \\ 
0 & -\sin\theta & \cos\theta &0 \\
0 & 0 & 0 & 1
\end{array} \right)
\left( \begin{array}{c}
 0 \\ 0 \\ 1 \\0
\end{array} \right)
= {1 \over \sqrt{2}}
\left( \begin{array}{c}
 0 \\ \sin\theta \\ \cos\theta \\0
\end{array} \right)
 \label{eqn2}
\end{eqnarray}
and for $\theta = 90$ degrees, $e^3_{\mu}$ is transformed into $e^2_{\mu}$.

However, there are good reasons to looks at the polarization states of composite particles. The author has suggested~\cite{perkins1} that all integral spin particles are composite particles formed of fermions. Also Varlamov~\cite{varlamov} has shown recently that the problems encountered in quantizing the electromagnetic field disappear if the photon is composed of fermion pairs. Thus, it seems very reasonable to consider four-vector fields constructed from fermions.

\section{Polarization States of Composite Particles}
\label{sec.polar}

Now let us look at the helicity states of a composite particle formed of a fermion-antifermion pair, each particle with mass $m$. Using the function $\Lambda (helicity)$ and up and down arrows to denote the z-component of spin for the spins ${1 \over 2}$ particles, we have the usual four combinations with well-defined exchange symmetry,

\newpage
\begin{eqnarray}
\Lambda(1) = \uparrow \uparrow \nonumber \\
\Lambda(-1) = \downarrow \downarrow \nonumber \\
\Lambda(0) = {1 \over \sqrt{2}}( \uparrow \downarrow + \downarrow \uparrow) \nonumber \\
\Lambda(0) = {1 \over \sqrt{2}}( \uparrow \downarrow - \downarrow \uparrow)
\label{eqn3}
\end{eqnarray}

To determine the polarization vectors for the different combinations of (\ref{eqn3}), we consider the vector field created from the fermion-antifermion pair,
\begin{equation}     
\Psi^\dagger i \gamma_4 \gamma_{\mu} \Psi. 
\label{eqn3.02}
\end{equation} 
 
This assumes that the pair are bound by a local interaction that does not involve some quanta~\cite{fermi-yang}. Unlike the pseudoscalar interaction, for example, this vector interaction has the required property of being attractive between particle and antiparticle and repulsive between 
particles~\cite{nambu,fermi-yang}. The Feynman diagram for this interaction is chain-like as shown in Fig.~\ref{f2}. Nambu and Jona-Lasinio~\cite{nambu} have used the Bethe-Salpeter equation in the chain approximation to form a pseudoscalar particle as a bound fermion-antifermion pair, and Ohanian~\cite{ohanian} has used the Bethe-Salpeter equation in the chain approximation to form gravitons from bound fermion-antifermion pairs.


\epsfxsize=5.0in \epsfbox{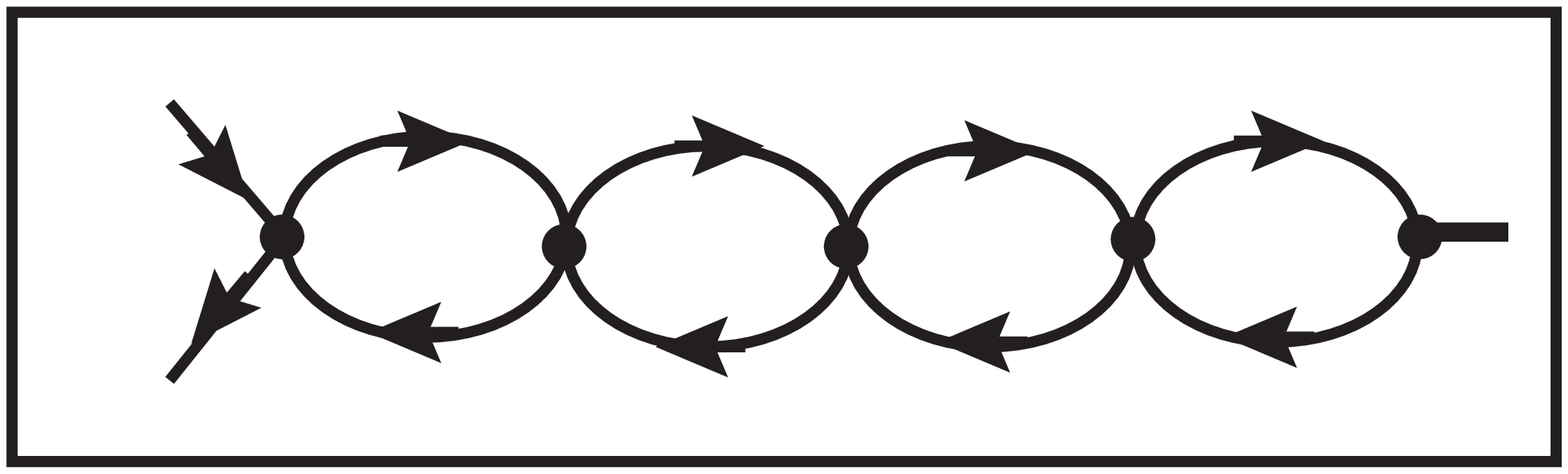}

\begin{figure}
\caption{Bound states as a chain of constituent fermion-antifermion bubbles.} \label{f2}
\end{figure}

Following Kane~\cite{kane} we can derive an important relation about the helicity of $\Psi^\dagger i \gamma_4 \gamma_{\mu} \Psi$. First we separate the upper and lower parts of the wave function,
\begin{eqnarray}     
\Psi =  \left( \begin{array}{c}     
\Psi_R \\
\Psi_L
\end{array} \right),
\label{eqn3.04}
\end{eqnarray} 
where L represents a left-handed positive energy solution and R represents a right-handed one.

We will use Weyl realization of the $\gamma$ matricies with, 
\begin{eqnarray}
\roarrow{\gamma} = i \left( \begin{array}{cc}
0 & \roarrow{\sigma} \\
-\roarrow{\sigma} & 0 
\end{array} \right), \;
 \gamma_4 = \left( \begin{array}{cc}
0 & 1 \\
1 & 0 
\end{array} \right), \;
\gamma_5 = \left( \begin{array}{cc}
-1 & 0 \\
0 & 1 
\end{array} \right). 
\label{eqn4.5}
\end{eqnarray}
where the ${\bf \sigma}$'s are the Pauli spin matricies.

\newpage
We can define projection operators,
\begin{eqnarray}
P_L = {{1 + \gamma_5} \over 2} = 
\left( \begin{array}{cc}
0 & 0 \\
0 & 1 
\end{array} \right), \;
P_R = {{1 - \gamma_5} \over 2} = 
\left( \begin{array}{cc}
1 & 0 \\
0 & 0 
\end{array} \right),
\label{eqn4.52}
\end{eqnarray}
whose effect on $\Psi$ is,
\begin{eqnarray}     
\Psi_L = P_L \Psi, \nonumber \\
\Psi_R = P_R \Psi. 
\label{eqn4.56}
\end{eqnarray} 

We can write,
\begin{eqnarray}     
\Psi^\dagger \gamma_4 \gamma_{\mu} \Psi = 
\Psi^\dagger \gamma_4 ( P_L + P_R) \gamma_{\mu} ( P_L + P_R) \Psi \nonumber \\
= \Psi^\dagger \gamma_4  P_L \gamma_{\mu}  P_L \Psi 
+\Psi^\dagger \gamma_4  P_R \gamma_{\mu}  P_L \Psi
+\Psi^\dagger \gamma_4  P_L \gamma_{\mu}  P_R \Psi
+\Psi^\dagger \gamma_4  P_R \gamma_{\mu}  P_R \Psi.
\label{eqn4.60}
\end{eqnarray} 
Since 
\begin{eqnarray}     
P_L \gamma_{\mu} = \gamma_{\mu} P_R, \nonumber \\
P_R \gamma_{\mu} = \gamma_{\mu} P_L,
\label{eqn4.62}
\end{eqnarray} 
the first and fourth terms of Eq.~(\ref{eqn4.60}) are zero and we also have,
\begin{eqnarray}     
\Psi^\dagger_L \gamma_4 = \Psi^\dagger \gamma_4 P_R, \nonumber \\
\Psi^\dagger_R \gamma_4 = \Psi^\dagger \gamma_4 P_L.
\label{eqn4.64}
\end{eqnarray} 

Substituting Eq.~(\ref{eqn4.64}) into Eq.~(\ref{eqn4.60}), we obtain,
\begin{equation}     
\Psi^\dagger \gamma_4 \gamma_{\mu} \Psi 
= \Psi^\dagger_L \gamma_4 \gamma_{\mu} \Psi_L
+ \Psi^\dagger_R \gamma_4 \gamma_{\mu} \Psi_R,
\label{eqn4.68}
\end{equation} 
which shows that the helicity is preserved for an interaction of the form
$\Psi^\dagger i \gamma_4 \gamma_{\mu} \Psi$.

For determining the polarization vectors, 
we will separate out the space-time dependence by putting 
$\Psi = u e^{i( {\bf p} \cdot {\bf x} - E t )}$ and use plane wave spinors that are solutions of the Dirac equation, 

\begin{eqnarray}
u^{+1}_{+1}({\bf p}) = \sqrt{ {E + p_3} \over 2 E} 
\left( \begin{array}{c}
1 \\ {{p_1 + i p_2} \over {E + p_3}} \\
{m \over {E + p_3}} \\ 0
\end{array} \right), 
\nonumber \\
u^{-1}_{-1}({\bf p}) = \sqrt{ {E + p_3} \over 2 E} 
\left( \begin{array}{c}
{{-p_1 + i p_2} \over {E + p_3}} \\ 1 \\
0 \\ {-m \over {E + p_3}} 
\end{array} \right), 
\nonumber \\
u^{-1}_{+1}({\bf p}) = \sqrt{ {E + p_3} \over 2 E} 
\left( \begin{array}{c}
{-m \over {E + p_3}} \\ 0 \\
1 \\ {{p_1 + i p_2} \over {E + p_3}} \\
\end{array} \right),
\nonumber \\
u^{+1}_{-1}({\bf p}) = \sqrt{ {E + p_3} \over 2 E} 
\left( \begin{array}{c}
0 \\ {m \over {E + p_3}} \\  {{-p_1 + i p_2} \over {E + p_3}} \\ 1
\end{array} \right), 
\label{eqn5}
\end{eqnarray}
where $p_\mu = ({\bf p},iE)$, and the superscripts and subscripts on $u$ refer to the energy and helicity states respectively.

Taking the composite particle to be at rest, one fermion will have momentum ${\bf p}$ and the other momentum ${\bf -p}$. The spinors for negative momenta are related to those of positive momenta by,

\begin{eqnarray}
u^{+1}_{+1}({\bf -p}) = u^{-1}_{-1}({\bf p}),  \nonumber \\
u^{-1}_{-1}({\bf -p}) = u^{+1}_{+1}({\bf p}),  \nonumber \\
u^{+1}_{-1}({\bf -p}) = u^{-1}_{+1}({\bf p}),  \nonumber \\
u^{-1}_{+1}({\bf -p}) = u^{+1}_{-1}({\bf p}).
\label{eqn6}
\end{eqnarray}

The polarization vectors corresponding to the different combinations of (\ref{eqn3}) are,
\begin{eqnarray}
\epsilon_\mu^1( p ) = {1 \over \sqrt{2}} [u^{-1}_{-1}({\bf p})]^\dagger i \gamma_4 \gamma_{\mu} u^{+1}_{+1}({\bf p}),  \nonumber \\
\epsilon_\mu^2( p ) = {1 \over \sqrt{2}} [u^{+1}_{+1}({\bf p})]^\dagger i \gamma_4 \gamma_{\mu} u^{-1}_{-1}({\bf p}),  \nonumber \\
\epsilon_\mu^3( p ) = {1 \over 2 \sqrt{2}} \left( [u^{+1}_{+1}({\bf p})]^\dagger i \gamma_4 \gamma_{\mu} u^{+1}_{+1}({\bf p}) + [u^{+1}_{-1}({\bf p})]^\dagger i \gamma_4 \gamma_{\mu} u^{+1}_{-1}({\bf p}) \right), \nonumber \\
\epsilon_\mu^4( p ) = {1 \over 2 \sqrt{2}} \left( [u^{+1}_{+1}({\bf p})]^\dagger i \gamma_4 \gamma_{\mu} u^{+1}_{+1}({\bf p}) - [u^{+1}_{-1}({\bf p})]^\dagger i \gamma_4 \gamma_{\mu} u^{+1}_{-1}({\bf p}) \right).
\label{eqn7}
\end{eqnarray}

Carrying out the matrix multiplications results in,
\begin{eqnarray}     
\epsilon_\mu^1(p) \!= \!{1 \over \sqrt{2}} \left( 
{{-i p_1 p_2 \!+\!E^2 \!+\!p_3 E \!-\!p_1^2} \over {E(E + p_3)}},
{{- p_1 p_2 \! + \!iE^2 \! +\!ip_3 E \! - \!ip_2^2 } 
\over {E(E + p_3)}},
{{\!-p_1 \!- \!i p_2} \over E}, 0 \right), \nonumber\\ 
\epsilon_\mu^2(p) \!= \!{1 \over \sqrt{2}} \left( 
{{i p_1 p_2 \!+\!E^2 \!+\!p_3 E \!-\!p_1^2} \over {E(E + p_3)}},
{{-p_1 p_2 \! - \!iE^2 \! -\!ip_3 E \! + \!ip_2^2 } 
\over {E(E + p_3)}},
{{\!-p_1 \!+ \!i p_2} \over E}, 0 \right), \nonumber\\ 
\epsilon_\mu^3(p) \!= \!{1 \over \sqrt{2}} \left( 
{p_1 \over E}, {p_2 \over E}, {p_3 \over E}, i \right), \nonumber\\ 
\epsilon_\mu^4(p) \!= \!\left( 0,0,0,0 \right).
\label{eqn8}
\end{eqnarray}     
As expected $\epsilon_\mu^4(p)$ vanishes, showing that there is no vector state for the last combination of Eq.~(\ref{eqn3}). 

There are several problems with these polarization vectors. First of all, there is the normalization problem,
\begin{eqnarray}     
\epsilon_\mu^1(p) \cdot \epsilon_\mu^{1*}(p)  \!= \!  1 +
 {{m^2 (m^2 \!+\! p_3^2 \! - \!E^2)} \over {2 E^2(E + p_3)^2}}, \nonumber\\ 
\epsilon_\mu^2(p) \cdot \epsilon_\mu^{2*}(p)  \!= \! 1 + {{m^2  
(m^2 \!+\! p_3^2 \! - \!E^2)} \over {2 E^2(E + p_3)^2}}, \nonumber\\ 
\epsilon_\mu^3(p) \cdot \epsilon_\mu^{3*}(p)  \!= \! 1 - {m^2 \over (2 E^2)},
\label{eqn9}
\end{eqnarray}     
instead of $\epsilon_\mu^j(p) \cdot \epsilon_\mu^{j*}(p) = 1$. Another problem is that the different polarization vectors are not orthogonal,
\begin{eqnarray}     
\epsilon_\mu^1(p) \cdot \epsilon_\mu^{2*}(p)  \!= \!  
 {{m^2 (-m^2 \!-\! 2 i p_1 p_2 \! - 2 p_1^2 - 2 p_3^2 + \!E^2)} 
\over {2 E^2(E + p_3)^2}}, \nonumber\\ 
\epsilon_\mu^1(p) \cdot \epsilon_\mu^{3*}(p)  \!= \! 
{{m^2  (p_1 \! + \! i p_2)} \over {2 E^2(E + p_3)}}, \nonumber\\ 
\epsilon_\mu^2(p) \cdot \epsilon_\mu^{3*}(p)  \!= \! 
{{m^2  (p_1 \! - \! i p_2)} \over {2 E^2(E + p_3)}}, 
\label{eqn10}
\end{eqnarray}     
instead of $\epsilon_\mu^j(p) \cdot \epsilon_\mu^{k*}(p) = 0$ for $j \ne k$. Furthermore, the dot product of each polarization vector with the internal four-momentum $p_\mu$ is not equal to zero.

All of these problems can be overcome immediately if we assume that the massive composite particle is formed of two {\it massless} fermions (i.e., set $m = 0$).
The composite particle could obtain mass as in the Nambu-Jona-Lasino dynamic model~\cite{nambu} or by the Higgs mechanism~\cite{kane}. 
If the fermions are massless, the polarization vectors depend only upon the direction of ${\bf p}$ (the internal momentum), ${\bf n} = {\bf p}/ |{\bf p}|$. Equation (\ref{eqn8}) now becomes,

\begin{eqnarray}     
\epsilon_\mu^1(n) \!= \!{1 \over \sqrt{2}} \left( 
{{-i n_1 n_2 \!+\!1 \!+\!n_3 \!-\!n_1^2} \over {1 + n_3}},
{{- n_1 n_2 \!+ \!in_1^2 \!+ \!in_3^2 \!+ \!in_3} 
\over {1 + n_3}},
\!-n_1 \!- \!i n_2, 0 \right), \nonumber\\ 
\epsilon_\mu^2(n) \!= \!{1 \over \sqrt{2}}\left( 
{{i n_1 n_2 \!+\!1 \!+\!n_3 \!-\!n_1^2} \over {1 + n_3}},
{{- n_1 n_2 \!- \!in_1^2 \!- \!in_3^2 \!- \!in_3} 
\over {1 + n_3}},
\!-n_1 \!+ \!i n_2, 0 \right), \nonumber\\
\epsilon_\mu^3(n) \!= \!{1 \over \sqrt{2}}\left( 
n_1, n_2, n_3, i \right).  
\label{eqn11}
\end{eqnarray}     
These polarization vectors satisfy the 
normalization relation,

\begin{eqnarray}     
\epsilon_\mu^j(n) \cdot \epsilon_\mu^{j*}(n) = 1, \nonumber\\
\epsilon_\mu^j(n) \cdot \epsilon_\mu^{k*}(n) = 0 \;\; \text{for} \;\; k \ne j.
\label{eqn11a}
\end{eqnarray}     

The Lorentz-invariant dot 
products of the internal four-momentum $p_\mu = |{\bf p}|(n_1,n_2,n_3,i)$
with the polarization vectors are,
\begin{eqnarray}     
p_\mu \epsilon_\mu^1(n) = 0, \nonumber\\ 
p_\mu \epsilon_\mu^2(n) = 0, \nonumber\\ 
p_\mu \epsilon_\mu^3(n) = 0.
\label{eqn12}
\end{eqnarray}     

The helicity operator acting upon the polarization vectors gives,

\begin{eqnarray}     
({\bf J \cdot n}) \epsilon_\mu^1(n) = \epsilon_\mu^1(n), \nonumber\\ 
({\bf J \cdot n}) \epsilon_\mu^2(n) = -\epsilon_\mu^2(n), \nonumber\\ 
({\bf J \cdot n}) \epsilon_\mu^3(n) = 0,
\label{eqn13}
\end{eqnarray} 

where the components of ${\bf J}$ are,

 \begin{eqnarray}
J_1 = \left( \begin{array}{cccc}
0 & 0 & 0 & 0 \\
0 & 0& -i & 0 \\ 
0 & i & 0 & 0 \\
0 & 0 & 0 & 0
\end{array} \right),
\; \; \; \; J_2 = \left( \begin{array}{cccc}
0 & 0 & i & 0 \\
0 & 0& 0& 0 \\ 
-i & 0& 0 & 0 \\
0 & 0 & 0 & 0
\end{array} \right),
\; \; \; \; J_3 = \left( \begin{array}{cccc}
0 & -i & 0 & 0 \\
i & 0 & 0& 0 \\ 
0 & 0& 0 & 0 \\
0 & 0 & 0 & 0
\end{array} \right),
\label{eqn14}
\end{eqnarray}
   
Under a rotation about ${\bf n}$ by an angle $\theta$, they change as follows,
\begin{eqnarray}     
\epsilon_\mu^1(n) \rightarrow e^{i\theta} \epsilon_\mu^1(n), \nonumber\\ 
\epsilon_\mu^2(n) \rightarrow e^{-i\theta} \epsilon_\mu^2(n), \nonumber\\ 
\epsilon_\mu^3(n) \rightarrow \epsilon_\mu^3(n).
\label{eqn15}
\end{eqnarray}     
Thus, $\epsilon_\mu^1(n)$, $\epsilon_\mu^2(n)$, and $\epsilon_\mu^3(n)$ correspond to states with helicity  $+1$, $-1$, and $0$ respectively.   

If the momentum 
is along the third axis, the
polarization vectors reduce to,
\begin{eqnarray}     
\epsilon_\mu^1(n) = {1 \over \sqrt{2}}(1,i,0,0), \nonumber\\ 
\epsilon_\mu^2(n) = {1 \over \sqrt{2}}(1,-i,0,0), \nonumber\\ 
\epsilon_\mu^3(n) = {1 \over \sqrt{2}}(0,0,1,i).  
\label{eqn16}
\end{eqnarray} 
    
Thus, $\epsilon_\mu^1(n)$ and $\epsilon_\mu^2(n)$ reduce to the usual polarization vectors for right and left circular-polarized photons respectively
while $\epsilon_\mu^3(n)$ corresponds to longitudinal polarization. The polarization vectors cannot be transformed into each other under a Lorentz transformation because they are orthogonal (see Eq.~\ref{eqn11a}). 

It has been noted that one could consider each transverse polarization as an independent particle (see Ref.~\cite{veltman}, p. 173, 180-2). This is because these polarization vectors are independent degrees of freedom and under a Lorentz transformation change into themselves. (For invariance under parity, one needs both transverse polarizations.) The longitudinal polarization causes problems if one considers it to be part of the photon field~\cite{evans}. We will take this longitudinal polarization as belonging to an independent particle with helicity-0. One does not usually think of the longitudinal polarization of a massive particle as being invariant, but as shown in Sec.~\ref{sec.polar} the helicity is preserved for an interaction of the form
$\Psi^\dagger i \gamma_4 \gamma_{\mu} \Psi$.

 As discussed in Sec.~\ref{sec.intro}, if one can impose the Lorentz condition on a four-vector field, this reduces the spin possibilities to spin-1 and eliminates spin-0. We will now apply the Lorentz condition to the above four-vector fields. The photon field can be constructed~\cite{perkins2} from $\epsilon_\mu^1(n)$ and $\epsilon_\mu^2(n)$,
\begin{eqnarray}
A_\mu(R) =  \sum_{\bf P} {1 \over \sqrt{2 V \omega_p}}\left\{ 
\left[\gamma_R({\bf P})\epsilon_\mu^1({\bf n})
+ \gamma_L({\bf P})\epsilon_\mu^2({\bf n}) 
\right]e^{i P R} \right. \nonumber \\
\left. + \left[\gamma_R^\dagger({\bf P})\epsilon_\mu^{1*}({\bf n})
+ \gamma_L^\dagger({\bf P})\epsilon_\mu^{2*}({\bf n}) 
\right]e^{-i P R}  \right\},
\label{eqn17}
\end{eqnarray}

This four-vector field satisfies the Lorentz condition,
\begin{equation}
(\partial A_\mu / \partial x_\mu) = 0
\label{eqn18}
\end{equation}
which follows from (\ref{eqn12}). 

\newpage
In a similar manner the field $V_\mu(R)$ for the particle with longitudinal polarization is,
\begin{eqnarray}
V_\mu(R) =  \sum_{\bf P} {1 \over \sqrt{2 V \omega_p}}\epsilon_\mu^3({\bf n})
\left\{ 
\left[\chi({\bf P}) - \eta({\bf P})
\right]e^{i P R} \right. \nonumber \\
\left. + \left[\chi^\dagger({\bf P})
- \eta^\dagger({\bf P}) 
\right]e^{-i P R}  \right\},
\label{eqn19}
\end{eqnarray}
with $\chi({\bf P})$ and $\eta({\bf P})$ being annihilation operators for the composite particle. Here $P R = {\bf P} \cdot {\bf R} - \sqrt{ {\bf P}^2 + M^2} t$ for a particle with mass $M$ and momentum ${\bf P}$. The vector ${\bf n}$ refers to the internal momentum direction and not ${\bf P}$. Since in the rest frame $P_\mu = (0,0,0,iM)$, we now have,
\begin{equation}     
P_\mu \epsilon_\mu^3(n) = -M.
\label{eqn20}
\end{equation}     
The field equations for this particle are,
\begin{eqnarray}     
(\Box - M^2) V_\mu = 0, \nonumber\\ 
(\partial V_\mu / \partial x_\mu) = -M f_V.
\label{eqn21}
\end{eqnarray}  
where   
\begin{eqnarray}
f_V(R) =  i\sum_{\bf P} {1 \over \sqrt{2 V \omega_p}}
\left\{ 
\left[\chi({\bf P}) - \eta({\bf P})
\right]e^{i P R} \right. \nonumber \\
\left. - \left[\chi^\dagger({\bf P})
- \eta^\dagger({\bf P}) 
\right]e^{-i P R}  \right\}.
\label{eqn22}
\end{eqnarray}
From the last equation of (\ref{eqn21}) we see that this four-vector field does {\it not} satisfy the Lorentz condition. Note that the composite-particle 
four-vector field of Eq.~(\ref{eqn19}) is not the four-gradient of a scalar field, such as that discussed in~\cite{weinberg}, since the polarization vectors are functions of ${\bf n}$, not ${\bf P}$.

\section{Conclusion }
\label{sec.concl}
                                   
The conditions under which a massive vector particle could exist in a single helicity state have been examined. The requirements are that the particle must be composed of a massless fermion-antifermion pair, which are bound by an interaction of the form 
$\Psi^\dagger i \gamma_4 \gamma_{\mu} \Psi$. 

If a massive helicity-0 vector particle exists, it will have characteristics similar to those of a pseudoscalar. However, since it has an axis with a direction, a helicity-0 particle could be identified experimentally by detecting an asymmetry in its decay products. The longitudinal polarization can lead to 
forward-backward asymmetries. Unlike a charged particle with spin $1/2$ or spin $1$, a charged helicity-0 particle will have zero magnetic moment and its direction of polarization will not be altered by a magnetic field.

It has been noted~\cite{lee-yang} that nature uses simple but odd constructions. The longitudinally polarized vector field, constructed in Sec.~\ref{sec.polar} is about the simplest vector field imaginable. The Proca model~\cite{ryder} of a spin-1 vector field for a massive particle with transverse and longitudinal polarizations is obviously much more complex.                                   

\acknowledgments
 
Many helpful discussions with Prof. J. E. Kiskis are 
gratefully acknowledged.

\end{document}